\documentclass[12pt]{article}
\usepackage{graphicx}  % needed for figures
\usepackage{subfigure}
\usepackage{dcolumn}   % needed for some tables
\usepackage{bm}        % for math
\usepackage{amssymb}   % for math
\usepackage{amsmath}   % for math
\usepackage{ulem}
\usepackage{dcolumn}
\usepackage{epsfig}
\usepackage{array}
\usepackage{url}
\newcommand{\gev}  {\mbox{${\rm GeV}$}}

\newcommand{\tev}  {\mbox{${\rm TeV}$}}

\newcommand{\invfb}{\mbox{${\rm fb}^{-1}$}}

\newcommand{\lum}  {\mbox{${\cal L}$}}

\newcommand{\Br}  {\mbox{${\cal B}$}}   % Branching ratio

\newcommand{\etal} {\mbox{$et~al.$}}

\newcommand{\pt}  {\mbox{$p_{\rm T}$}}
\newcommand{\ptvis}  {\mbox{$p_{\rm T}^{\rm vis}$}}
\newcommand{\et}  {\mbox{$E_{\rm T}$}}

\newcommand{\met} {\mbox{${E\!\!\!\!/_{\rm T}}$}}

\newcommand{\mtt}{\mbox{$M_{\tau\tau}$}}

\newcommand{\mttend}{\mbox{$M_{\tau\tau}^{\rm end}$}}
\newcommand{\mjtt}{\mbox{$M_{j\tau\tau}$}}
\newcommand{\mjttpeak}{\mbox{$M_{j\tau\tau}^{\rm peak}$}}
\newcommand{\mjttsame}{\mbox{$M_{j\tau\tau}^{\rm same}$}}
\newcommand{\mjttbi}{\mbox{$M_{j\tau\tau}^{\rm bi^\_event}$}}
\newcommand{\mjt}{\mbox{$M_{j\tau}$}}
\newcommand{\mjtFirst}{\mbox{$M_{j\tau}^{(1\mathrm{st})}$}}
\newcommand{\mjtSecond}{\mbox{$M_{j\tau}^{(2\mathrm{nd})}$}}

\newcommand{\mjtend}{\mbox{$M_{j\tau}^{\rm end}$}}
\newcommand{\meff}{\mbox{$M_{{\rm eff}}$}}
\newcommand{\meffpeak}{\mbox{$M_{{\rm eff}}^{\rm peak}$}}

\newcommand{\meffbnoW}{\mbox{$M_{{\rm eff}}^{(b,~\mathrm{no}\ W)}$}}
\newcommand{\meffbnoWpeak}{\mbox{$M_{{\rm eff}}^{(b,~\mathrm{no}\ W)\; \rm peak}$}}

\newcommand{\mjW}{\mbox{$M_{jW}$}}
\newcommand{\mjWend}{\mbox{$M_{jW}^{\rm end}$}}
\newcommand{\mjj}{\mbox{$M_{jj}$}}
\newcommand{\mjjsame}{\mbox{$M_{jj}^{\rm same}$}}
\newcommand{\mjjbi}{\mbox{$M_{jj}^{\rm bi^\_event}$}}

\newcommand{\azero}{\mbox{$A_{0}$}}
\newcommand{\tanb} {\mbox{$\tan\beta$}}
\newcommand{\sinb} {\mbox{$\sin\beta$}}
\newcommand{\mzero}{\mbox{$m_{0}$}}
\newcommand{\mhalf}{\mbox{$m_{1/2}$}}

\newcommand{ \gluino}   {\mbox{$\tilde{g}$}}
\newcommand{ \squark}   {\mbox{$\tilde{q}$}}

\newcommand{ \usquarkL}  {\mbox{$\tilde{u}_{L}$}}
\newcommand{ \usquarkR}  {\mbox{$\tilde{u}_{R}$}}

\newcommand{ \sbottomone}{\mbox{$\tilde{b}_{1}$}}

\newcommand{ \sbottomtwo}{\mbox{$\tilde{b}_{2}$}}

\newcommand{ \stopone}  {\mbox{$\tilde{t}_{1}$}}

\newcommand{ \stoptwo}  {\mbox{$\tilde{t}_{2}$}}

\newcommand{ \seleR}    {\mbox{$\tilde{e}_{R}$}}

\newcommand{ \seleL}    {\mbox{$\tilde{e}_{L}$}}

\newcommand{ \stauone}  {\mbox{$\tilde{\tau}_{1}$}}

\newcommand{ \stauonepm}{\mbox{$\tilde{\tau}_{1}^{\pm}$}}
\newcommand{ \stautwo}  {\mbox{$\tilde{\tau}_{2}$}}

\newcommand{ \schionezero }{\mbox{$\tilde{\chi}_{1}^{0}$}}
\newcommand{ \schitwozero }{\mbox{$\tilde{\chi}_{2}^{0}$}}
\newcommand{ \schithreezero }{\mbox{$\tilde{\chi}_{3}^{0}$}}
\newcommand{ \schifourzero }{\mbox{$\tilde{\chi}_{4}^{0}$}}
\newcommand{ \schionepm }{\mbox{$\tilde{\chi}_{1}^{\pm}$}}
\newcommand{ \schionemp }{\mbox{$\tilde{\chi}_{1}^{\mp}$}}

\newcommand{ \schitwopm }{\mbox{$\tilde{\chi}_{2}^{\pm}$}}

\newcommand{ \pythia } {{\tt PYTHIA}}
\newcommand{ \isasugra } {{\tt ISASUGRA}}
\newcommand{ \isajet }    {{\tt ISAJET}}
\newcommand{ \pgs }    {{\tt PGS4}}
\newcommand{ \darksusy }    {{\tt darkSUSY}}
\newcommand{ \DMrelic }{\mbox{$\Omega_{\schionezero}h^{2}$}}

\newcommand{ \osls }{\mbox{OS$-$LS~}}

\makeatletter
\def\alt{\mathrel{\mathpalette\gl@align<}}
\def\agt{\mathrel{\mathpalette\gl@align>}}
\def\gl@align#1#2{\lower.6ex\vbox{\baselineskip\z@skip\lineskip\z@
\ialign{$\m@th#1\hfil##\hfil$\crcr#2\crcr\sim\crcr}}}
\makeatother

\begin{document}
\begin{flushright}
%{\tt }\\
MIFPA-10-37

\end{flushright}
\vspace*{2cm}
\begin{center}
{\baselineskip 25pt \large{\bf
Determination of Non-Universal Supergravity Models at the Large Hadron Collider
} \\

}

\vspace{1cm}

{\large
Bhaskar~Dutta$^{1}$, Teruki~Kamon$^{1,2,3}$, Abram~Krislock$^{1}$, Nikolay~Kolev$^{4}$, Youngdo~Oh$^{3}$
} \vspace{.5cm}

{
\it $^{1}$ Department of Physics, Texas A\&M University, College Station, TX 77843-4242, USA\\
$^{2}$ Fermi National Accelerator Laboratory, Batavia, Illinois 60510, USA\\
$^{3}$ Department of Physics, Kyungpook National University, Daegu 702-701, South Korea\\
$^{4}$ Department of Physics, University of Regina, Regina, SK S4S 0A2, Canada\\
}
\vspace{.5cm}

\vspace{1.5cm} {\bf Abstract}\end{center}

We examine a well motivated non-universal supergravity model where the Higgs boson masses are not unified with the other scalars at the grand unified scale at the LHC. The dark matter content can easily be satisfied in this model by having a larger Higgsino component in the lightest neutralino. Typical final states in such a scenario at the LHC involve $W$ bosons. We develop a bi-event subtraction technique to reduce a huge combinatorial background to identify $W\rightarrow jj$ decays. This is also a key technique to reconstruct supersymmetric particle masses in order to determine the model parameters. With the model parameters, we find that the dark matter content of the universe can be determined in agreement with existing experimental results.

\thispagestyle{empty}

\bigskip
\newpage

\addtocounter{page}{-1}

\section{Introduction}

The dark matter content of the universe today has been measured very precisely by the WMAP experiment~\cite{WMAP} which shows that the energy density of the universe is comprised of 23\% dark matter. Supersymmetry (SUSY) models with conservation of R-parity can naturally explain the dark matter content, as well as solving many problems inherent in the Standard Model (SM). In most SUSY models, the weakly interacting lightest neutalino %is neutral and stable, making it an 
is an excellent dark matter candidate~\cite{neuDM} since the dark matter content of the universe can be satisfied in these SUSY models. 

The existence of this dark matter connection is under tremendous experimental investigation at the Large Hadron Collider (LHC) and at direct and indirect dark matter detection experiments. At the LHC, the dark matter hypothesis can be tested by producing the dark matter particles which will give rise to missing energy signal. In addition to the dark matter candidate, other SUSY particles will also be produced. Attempts will be made to measure SUSY particle masses and model parameters which will be used to estimate the dark matter content based on the available measurements at the LHC. It will be very interesting if such an estimate of the dark matter relic density is close to the measurement of WMAP~\cite{WMAP} because we will be on the verge of establishing a true connection between particle physics and cosmology. Of course, the measurements from the direct and indirect detection experiments need to support this connection as well.

The Minimal Supersymmetric Standard Model (MSSM) is a very general supersymmetric extension of the SM which has more than a hundred parameters to specify the model. It would be impossible to completely reconstruct this entire model at the LHC since that would require finding more than a hundred measurable quantities to determine the model parameters. Extraction of the measurable quantities, end points, and peak positions of different kinematical distributions is not easy due to severe background problems from the model itself. \textquotedblleft Bottom-up\textquotedblright\ studies which aim at using measurements made at the LHC to reconstruct MSSM model parameters have been performed in the past~\cite{sps}. Both model-dependent (with assumptions about unifying parameters) and model-independent (more general MSSM) methods have been investigated in these studies. Our approach to the \textquotedblleft bottom-up\textquotedblright\ study of SUSY is to use well motivated SUSY models with fewer parameters to study and understand the final states and associated observables. This way, we can determine all the model parameters and calculate the dark matter content. Also, the techniques we develop to extract the measurable quantities can also be applied to models with larger sets of parameters.

We first used the minimal supergravity motivated model (mSUGRA)~\cite{sugra1}. The mSUGRA model has the feature that many SUSY masses unify at the grand unified scale (GUT scale). This feature makes the model very simple, requiring only four parameters and a sign to determine all of the SUSY particle spectrum. The choice of the sign can be motivated by the branching ratio, $\Br(b\rightarrow s\gamma)$~\cite{bsgamma}. Thus to reconstruct this model at the LHC requires only four distinct measurements.

In previous studies~\cite{LHCtwotau, LHCthreetau, LHCrelicdensity, sscprd}, we have developed methods to determine mSUGRA parameters at the LHC. We found that the dark matter content can be measured with an accuracy comparable to the WMAP measurement. We also found that different dark matter allowed regions have different smoking gun signals at the LHC in the mSUGRA model. 

Having all scalar masses unified at the GUT scale is perhaps too simple an assumption. On the other hand, determining many parameters at the LHC is also a very difficult task. This study takes the first step in a more general direction. We study a non-universal supergravity (nuSUGRA) model, where the Higgs masses no longer unify at the GUT scale with the other scalar SUSY particles. This Higgs sector non-universality is easily imagined because it is a completely different sector of matter from the SUSY partners of the quarks and leptons. This type of model has been studied extensively in the context of dark matter~\cite{nath1,nuhm}.

This nuSUGRA model has six parameters instead of four, with two new parameters for the two Higgs doublet masses. The difficulty in any of these studies is to have enough experimentally measurable observables to determine all of the model parameters. In this work, we investigate the decay chains and final states of the model to identify typical signals and to find these observables. This task is well worth the effort even if this nuSUGRA model is not true, since the measurement techniques we uncover can be applied to any other model with similar signals to measure relations between SUSY particle masses.

The SUSY measurements at the LHC involve cascade decays arising from the colored SUSY particles, the squarks and gluinos. SUSY models with R-parity are difficult to measure, since each event has its own background. This is because R-parity demands that SUSY particles always be produced in pairs. Thus, each event has two SUSY decay chains which can be a background to each other. In this paper, we demonstrate a technique to isolate only the decay chain we want to look at by effectively subtracting out the signal from the other chain. This technique is a method which eliminates a large amount of background from the signal we want to measure.

This subtraction technique combines with kinematic distributions to make enough measurements to fully reconstruct all the model parameters. We then use those parameters to determine the entire SUSY spectrum, as well as to determine the dark matter relic density of the universe.

The outline of this paper is as follows. In section 2, we describe in more detail the nature of the nuSUGRA model. In section 3, we describe the signals that would be seen from this model at the LHC, as well as the observables and measurement techniques needed to fully determine this model. In section 4, we compile all the measurement results together to determine the model parameters and estimate their statistical and systematic uncertainties. We conclude in section 5.
%, giving the main results, which include the determined model parameters and the calculated relic density of the universe, as well as the estimated uncertainties for each. [{\bf Needs adjusting... But can't think of the correct phrasing right now.}]

\section{nuSUGRA Model and Benchmark Point}

We first review the mSUGRA model in order to describe the nuSUGRA model. The mSUGRA model has the attractive feature that many of the SUSY particle masses are unified at the GUT scale. Thus, it needs only four parameters and a sign to specify the entire model. These parameters are:
\begin{itemize}
\item The unified scalar mass at the GUT scale, $\mzero$,
\item The unified gaugino mass at the GUT scale, $\mhalf$,
\item The trilinear coupling at the GUT scale, $\azero$,
\item The ratio of the vacuum expectation values of the two Higgs
doublets, $\tanb$, and
\item The sign of the Higgs bilinear coupling, $\mathrm{sign}(\mu)$.
\end{itemize}

Since these model parameters (along with the SM parameters) specify all of the masses and mixings for the SUSY particles in the model, they also determine whether or not this model predicts the correct amount of dark matter left in the universe today. If we assume a history of the universe where the dark matter particles, the neutralinos, were in thermal equilibrium in the universe at early times, then a large region of the mSUGRA parameter space actually predicts too much dark matter today. This is due to the neutralinos not being able to annihilate enough in the early universe. However, certain regions of mSUGRA parameter space allow for mechanisms where the neutralino annihilation cross-section is large enough during early times which lead to the correct amount of dark matter today~\cite{darkrv}.

For instance, the co-annihilation region has the characteristic feature that the stau particle has a mass very close to that of the neutralino. This allows for the neutralino to co-annihilate with the lightest stau particle in the early universe. This extra annihilation mechanism increases the total annihilation cross-section for the neutralino~\cite{stauco}. It is also possible to have stop coannihilation if $\azero$ has a large yet negative value~\cite{stopco}. Another region, called the $A$-funnel region, has the feature that the neutralino mass is very close to being half the mass of the pseudo-scalar Higgs boson ($A^{0}$). Thus, there is a resonance when the neutralinos annihilate through this $A^{0}$ channel, which increases the annihilation cross-section~\cite{Afunnel}. Light Higgs resonance annihilation is also possible for small values of $m_{1/2}$~\cite{lighthiggs}.

A third region, called the focus point/hyperbolic branch region, has a \textquotedblleft focused\textquotedblright\ value for the parameter $\mu$~\cite{focuspoint}. This parameter is determined in mSUGRA by the electroweak symmetry breaking requirement. In the focus point region, this requirement causes the value of $\mu$ to be very small. The small value of $\mu$ causes the lightest neutralino to be very Higgs-like and couple strongly to heavier particles. Thus, neutralino annihilation diagrams containing $Z$ or Higgs bosons are favored. This effect causes the annihilation cross-section to be large enough to have the right amount of dark matter today.

There is another way to achieve this small value for the $\mu$ parameter: The nuSUGRA model. In the nuSUGRA model, the Higgs bosons are given a non-universal mass. Normally, for the mSUGRA model, the Higgs bosons, being scalar particles, have a mass of $\mzero$ at the GUT scale. Since the Higgs masses are intimately related to the electroweak symmetry breaking condition, adjusting the Higgs masses has a direct effect on the value of $\mu$. In a sense, we are promoting $\mu$ to a free parameter, since for any particular choice of the other four parameters, we can adjust $\mu$ by adjusting Higgs masses.

Since there are two Higgs doublets in SUSY models, we can have a parameter for each of their masses at the GUT scale, i.e., $m_{H_u}^2=(1+\delta_{H_u}) m_0^2$, $m_{H_d}^2=(1+\delta_{H_d}) m_0^2$. However, only one of the Higgs masses affects the value of $\mu$. The value of $\mu^2$ at the electroweak scale in terms of the GUT scale parameters is determined by the renormalization group equations (RGEs). In general, one must solve these numerically. However, one can get a qualitative understanding of the effects of the $\delta_H$'s from an analytic solution which is valid for low and intermediate $\tanb$~\cite{nath1}: 
\begin{equation} %%%%% Equation
\mu^2=\frac{t^2}{t^2-1} \left[
	\left(\frac{1-3 D_0}{2}-\frac{1}{t^2}\right)
	+\left(-\frac{1+D_0}{2}\delta_{H_u}+\frac{\delta_{H_d}}{t^2}\right)
\right]m_0^2 + \Delta,
\label{eqMu}
\end{equation} %%%%%
where $t\equiv \tanb$, $D_0\simeq 1-(m_t/200\ \sinb)^2$, and $\Delta$ contains the universal parts (which are independent of the $\delta_H$'s) and loop corrections. In general $D_0$ is small ($D_0\lesssim 0.23$). Equation \ref{eqMu} shows that $\mu$ is primarily sensitive to $\delta_{H_u}$. However, the pseudoscalar and heavy Higgs boson masses depend on both $\delta_{H_u}$ and $\delta_{H_d}$. 

In this model, the dark matter content can be satisfied not only by lowering $\mu$ but also having the pseuoscalar or heavy Higgs mass equal to twice the neutralino mass. Since we have two new parameters in the Higgs sector, both the pseudoscalar mass and $\mu$ are free parameters in this model. In the case where the dark matter content is satisfied by the heavy Higgs/pseudoscalar Higgs resonance, the heavy Higgs mass needs to be measured to see whether its mass obeys the resonant funnel condition. In our case we do not consider the Higgs funnel region but consider the first scenario where $\mu$ is changed to satisfy the dark matter content. For the purposes of this study, we choose one such model which predicted a dark matter relic density in agreement with that measured by WMAP. This scenario is also very interesting since it has large direct detection spin-independent cross-section of $3.56\times10^{-8}~\mathrm{pb}$ for proton collisions, and therefore it will be detected in the ongoing/upcoming runs of direct detection experiments~\cite{directDetection}.

Since $\mu$ is affected by only the up type Higgs, we define the nuSUGRA model with the unified Higgs mass at the GUT scale, $m_{H_u} = m_{H_d} \equiv m_H$, which becomes the fifth parameter of the model. The mass spectrum for our benchmark point of the nuSUGRA model is shown in Table~\ref{tabSpectrum}. We determine the mass spectrum for this model using \isasugra~\cite{isajet}. 

\begin{table}
\caption{SUSY masses and parameters (in $\gev$) for the point $\mzero = 360~\gev$,
$\mhalf = 500~\gev$, $\tanb = 40$, $\azero = 0$, and $m_H = 732~\gev$.  The top mass is set as $172.6~\gev$.
For this point, the dark matter relic density is $\DMrelic = 0.11$. The
total production cross-section for this point is $\sigma =
1.25~\mathrm{pb}$.}
 \label{tabSpectrum}
\begin{center}
\begin{tabular}{c c c c c c c c c c | c}
\hline \hline
$\gluino$ &
$\begin{array}{c} \usquarkL \\ \usquarkR \end{array}$ &
$\begin{array}{c} \stoptwo \\ \stopone \end{array}$ &
$\begin{array}{c} \sbottomtwo \\ \sbottomone \end{array}$ &
$\begin{array}{c} \seleL \\ \seleR \end{array}$ &
$\begin{array}{c} \stautwo \\ \stauone \end{array}$ &
$\begin{array}{c} \schitwozero \\ \schionezero \end{array}$ &
$\begin{array}{c} \schifourzero \\ \schithreezero \end{array}$ &
$\begin{array}{c} \schitwopm \\ \schionepm \end{array}$ &
$\begin{array}{c} A^0 \\ h^0 \end{array}$ &
$\mu$
\\ \hline
1161 &
$\begin{array}{c} 1113 \\ 1078 \end{array}$ &
$\begin{array}{c} 992 \\ 781 \end{array}$ &
$\begin{array}{c} 989 \\ 946 \end{array}$ &
$\begin{array}{c} 494 \\ 407 \end{array}$ &
$\begin{array}{c} 446 \\ 255 \end{array}$ &
$\begin{array}{c} 293 \\ 199 \end{array}$ &
$\begin{array}{c} 432 \\ 316 \end{array}$ &
$\begin{array}{c} 427 \\ 291 \end{array}$ &
$\begin{array}{c} 647 \\ 115 \end{array}$ &
307 
\\ \hline \hline
\end{tabular}
\end{center}
\end{table}

\section{Characteristic Signal and Observables at the LHC}

Our benchmark point of the nuSUGRA model shows some features which we observed in previous studies of mSUGRA scenarios~\cite{LHCtwotau, LHCthreetau, LHCrelicdensity, sscprd}. In particular, it still predicts that the LHC would see high $\pt$ jets from squark decays to neutralinos and charginos, many $\tau$'s from neutralino and stau decays, and large missing transverse energy ($\met$) from the lightest neutralino escaping the detector. However, in this nuSUGRA region, we also see another unique final state: There are many $W$ bosons being produced from neutralinos decaying into charginos or vice versa.

Since these $W$ bosons seem to be a smoking gun signal, we perform a random parameter space scan in mSUGRA to see if $W$ bosons are being produced or not. For values of $\tanb = \{10,40\}$, we scan $\mhalf < 1000~\gev$, and $\mzero < 2000~\gev$, keeping $\azero = 0$ and $\mu > 0$. We also ensure that experimental bounds for the Higgs mass~\cite{higgs1}, lightest chargino mass~\cite{aleph}, and squark masses~\cite{tevatron} are not violated. The results of this scan show that $W$ bosons in similar decay chains do appear in mSUGRA. However, under a thermal dark matter scenario, if we constrain the relic density of the universe to be somewhat close to the WMAP allowed region (using $\DMrelic < 0.3$), then we find that the only mSUGRA models with jet$+2\tau$ and jet+$W$ final states are in the co-annihilation region. This means that we can discern between mSUGRA and nuSUGRA at the LHC simply by measuring the difference in mass between the lightest stau and lightest neutralino
 (assuming a thermal dark matter scenario). For instance, in the coannihilation region of mSUGRA, the mass difference is $\Delta M_{\stauone - \schionezero} = 5$-$20~\gev$, whereas in our nuSUGRA benchmark point $\Delta M_{\stauone - \schionezero} = 56~\gev$.

Since this model can be discerned at the LHC, we study it to see if we can fully reconstruct the model. We use Monte Carlo programs which simulate a LHC experiment. To determine the mass spectrum of the model, we use \isasugra~\cite{isajet}. The mass spectrum is passed on to \pythia~\cite{pythia} to generate the Monte Carlo hard scattering events and hadron cascade. Unless otherwise specified, these events are $14~\tev$ $pp$ collisions. Each such event is then passed on to \pgs~\cite{pgs} to simulate the detector effects. 

Since we require five independent measurements to determine our five model parameters, we construct as many useful measurements as possible. To this end, it is necessary to utilize the $W$ boson decay chains. Reconstructing $W$ bosons from their hadronic decays is difficult in a jets plus $\met$ final state. The leptonic decays of $W$ decay are not adequate, since neutrinos, a source of $\met$, are also produced. We instead reconstruct $W$ bosons from their decays to quark pairs, i.e. jets in the detector. Thus, we must develop techniques of reconstructing $W$ bosons from these jets.

There are several mass reconstruction techniques for finding new particles, such as classical endpoint measurements~\cite{hinch1}, mass relation techniques~\cite{massRelation}, and $m_{{\rm T}2}$ techniques~\cite{mt2}. We choose the endpoint measurements as the simplest technique with minimum assumptions in this study to demonstrate our proposed subtraction technique. We also assume that the SM background shapes are studied and subtracted in order to focus on this technique as well as the observables we use to fully determine the nuSUGRA parameters. The subtraction technique and LHC simulations of the observables are described in the following subsections

\subsection{Bi-Event Subtraction Technique (BEST)}

One such technique aims at an ability to select the particular jet pair originating from $W$ boson decays. The difficulty is in discerning this jet pair from amongst the large combinatorial background of non-$W$ jets. We employ a subtraction technique to deal with this issue. This subtraction technique is similar to one used for a lepton plus jets system~\cite{bestOriginal}. We collect all the jet pairs in an event, each of which could be (a) both from the $W$ boson decay, (b) one from the $W$ boson and one from another source, or (c) both from non-$W$ sources. These are \textquotedblleft same-event\textquotedblright\ jet pairs with which we form the same-event dijet invariant mass distribution, $\mjjsame$. 

If we know the shape of the distribution containing (b) and (c) jet pairs, we can subtract this distribution from $\mjjsame$ to drastically reduce the background. To estimate the shape of this distribution, we collect all the jet pairs we can make by selecting one jet from the event of interest and one jet from a different event. We expect the \textquotedblleft bi-event\textquotedblright\ dijet invariant mass distribution, $\mjjbi$, to have a similar shape as most of the background in (b) and (c). Since there is no way for the bi-event jet pairs to come from a single $W$ boson, it definitely does not match the shape of the (a) distribution. We normalize the $\mjjbi$ distribution to the $\mjjsame$ distribution in the region of mass heavier than the $W$ boson. Then we subtract the $\mjjbi$ distribution from the $\mjjsame$ distribution to get $\mjj$. This final $\mjj$ distribution shows a nice $W$ boson mass peak with a much smaller background than the original $\mjjsame$ distribution. This \textquotedblleft bi-event subtraction technique\textquotedblright\ is the BEST we can do for finding $W$ bosons.

\begin{figure} [t!]
\centering
\includegraphics[width=.70\textwidth]{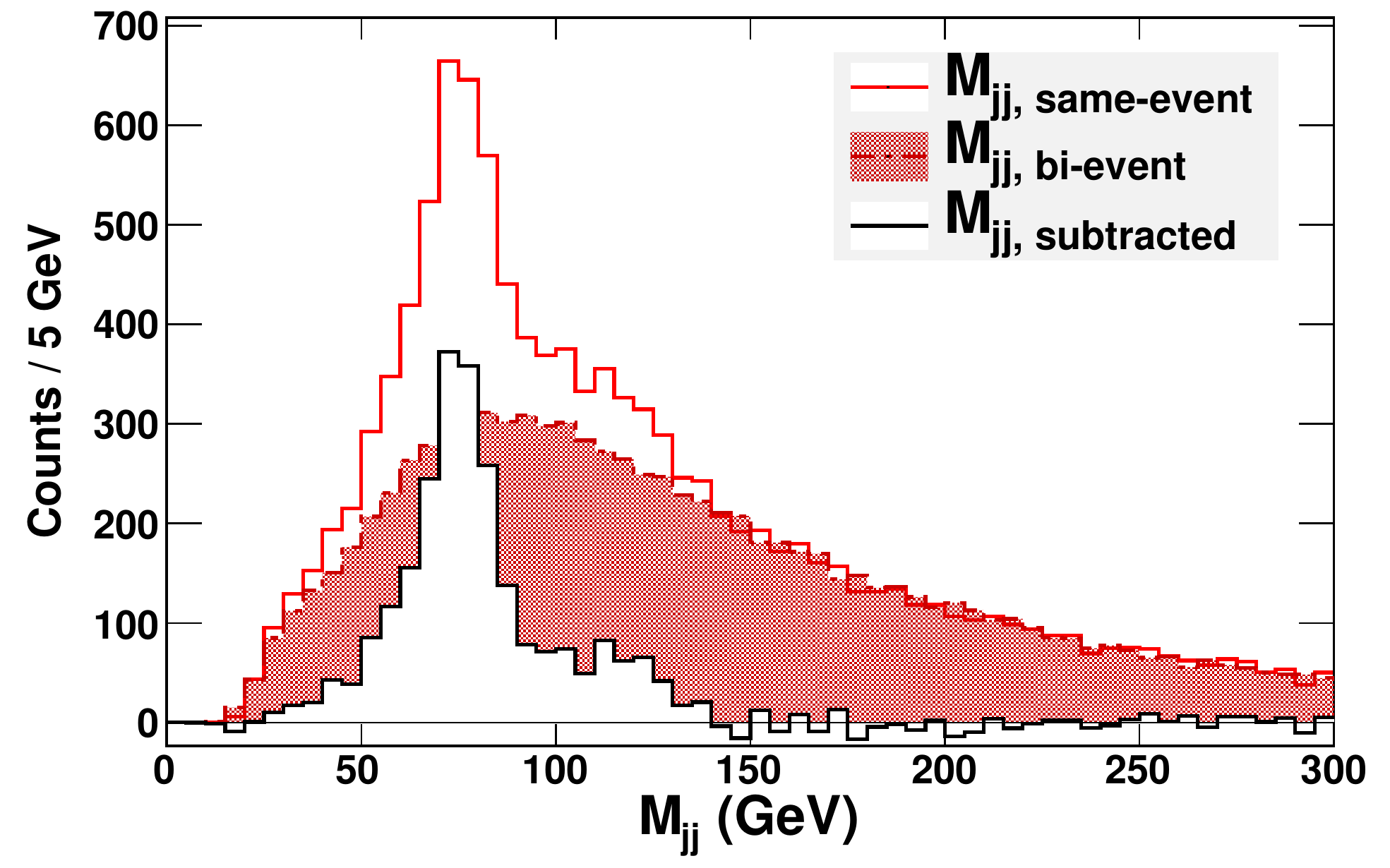}
\caption{The di-jet invariant mass distribution for $t\bar t$ in $7~\tev$ $pp$ collisions with a luminosity of $1~\invfb$. This figure demonstrates our BEST for finding $W$ bosons in the events. The solid red(grey) histogram is constructed using same-event jets. The dot-dashed and filled red(grey) histogram is constructed using jets from different (or bi-) events and is normalized to the shape of the long tail (above $200~\gev$) in the same-event histogram. The same-event minus bi-event subtraction produces the black subtracted histogram. We notice a clear reduction in the background shape around the $W$ mass peak.}
\label{figBEST}
\end{figure}

To demonstrate the BEST, we produced a small ($1~\invfb$) $t\bar t$ sample of $7~\tev$ $pp$ collisions. We selected events with the following cuts:
\begin{itemize}
\item Missing transverse energy, $\met \ge 25~\gev$;
\item Number of jets, $N_{\mathrm{jet}} \ge 4$ with jet $\pt \ge 25~\gev$ and $|\eta| \le 2.5$;
\item At least one of the four jets must be $b$ tagged;
\item Exactly one lepton ($e$ or $\mu$ with $\pt > 20~\gev$).
\end{itemize}
With these events, we find the $W$ mass peak without the top mass constraint in a $jjb$ final state. We select jet pairs with $\Delta R_{jj} \ge 0.4$, since jets which are too close cannot be discerned by the detector. The BEST $W$ finding for this sample is shown in Fig.~\ref{figBEST}.

%%%%% Figure
%%%%%

This technique helps us remove backgrounds from other signals as well. For instance, we often combine the two leading jets in each event (which come from the squark decays) with other reconstructed objects to form an observable. Doing our BEST helps us to choose the correct jet from the decay chain we want. We perform our BEST in almost all of the observables described below.

\subsection{$W$ Plus Jet}

The $W$ plus jet signal originates primarily from the following two decay chains:

\begin{subequations}   %%%%% Equation
  \label{decayJetW:list}
  \begin{align}
    \squark &\rightarrow q+\schionepm \rightarrow q+W^{\pm}+\schionezero  \label{decayJetW:Chargino} \\
    \squark &\rightarrow q+\schifourzero \rightarrow q+W^{\pm}+\schionemp \rightarrow \cdots  \label{decayJetW:Neutralino}
  \end{align}
\end{subequations}  %%%%%
The signal is characterized by a high $\pt$ jet and high $\met$ as well as a $W$ boson, which we see in the detector as two jets with invariant mass in the $W$ mass window ($65~\gev \le M_{jj} \le 90~\gev$). We reconstruct the $W$ boson and combine it with the corresponding leading $\pt$ jet from this decay chain to make the $W$ plus jet invariant mass, $\mjW$.

To select events for this signal, we use the following cuts:
\begin{itemize}
\item Missing transverse energy, $\met \ge 180~\gev$;
\item Number of jets, $N_{\mathrm{jet}} \ge 4$ with jet $\pt \ge 30~\gev$ and $|\eta| \le 2.5$. (Here we do not count $b$ tagged jets);
\item The two leading jets must have $\pt \ge 100~\gev$;
\item No leptons at all in the event (no $\tau$ with $\pt > 20~\gev$, and no $e$ or $\mu$ with $\pt > 5~\gev$);
\item The scalar sum, $\pt_{\mathrm{jet},1} + \pt_{\mathrm{jet},2} + \met \ge 600~\gev$;
\item There must be no $b$ tagged jet with $\pt$ larger than either of the two leading jets.
\end{itemize}
These cuts help to remove a lot of the SM background which will be seen at the LHC.  The dominant SM backgrounds for this process are $t\bar{t}$, $W$ plus jets, and $Z$ plus jets events.  After the cuts are performed, we begin to pair up all of the jets (except for the two leading jets) to look for $W$ candidates. The $W$ bosons which are produced by the decay chain in Eq.~\ref{decayJetW:list} can have large momentum from being near the bottom of the cascade decay. Thus, we expect the jets pairs from the $W$ to be close together due to Lorentz boosting. Again, the detector cannot discern the two jets if they are too close together. Therefore, we choose jet pairs which have $0.4 \le\Delta R_{jj}\le1.5$. We have used a similar cut in our previous studies~\cite{sscprd}.

We sort our jet pairs into two categories: Those that are in the $W$ mass window (where $65~\gev \le M_{jj} \le 90~\gev$), and those that are clearly not $W$'s which are in the sideband window (where $40~\gev \le M_{jj} \le 55~\gev$ or $100~\gev \le M_{jj} \le 115~\gev$). We sort them this way for an upcoming sideband subtraction.

\begin{figure} [t!]
\centering
\includegraphics[width=.70\textwidth]{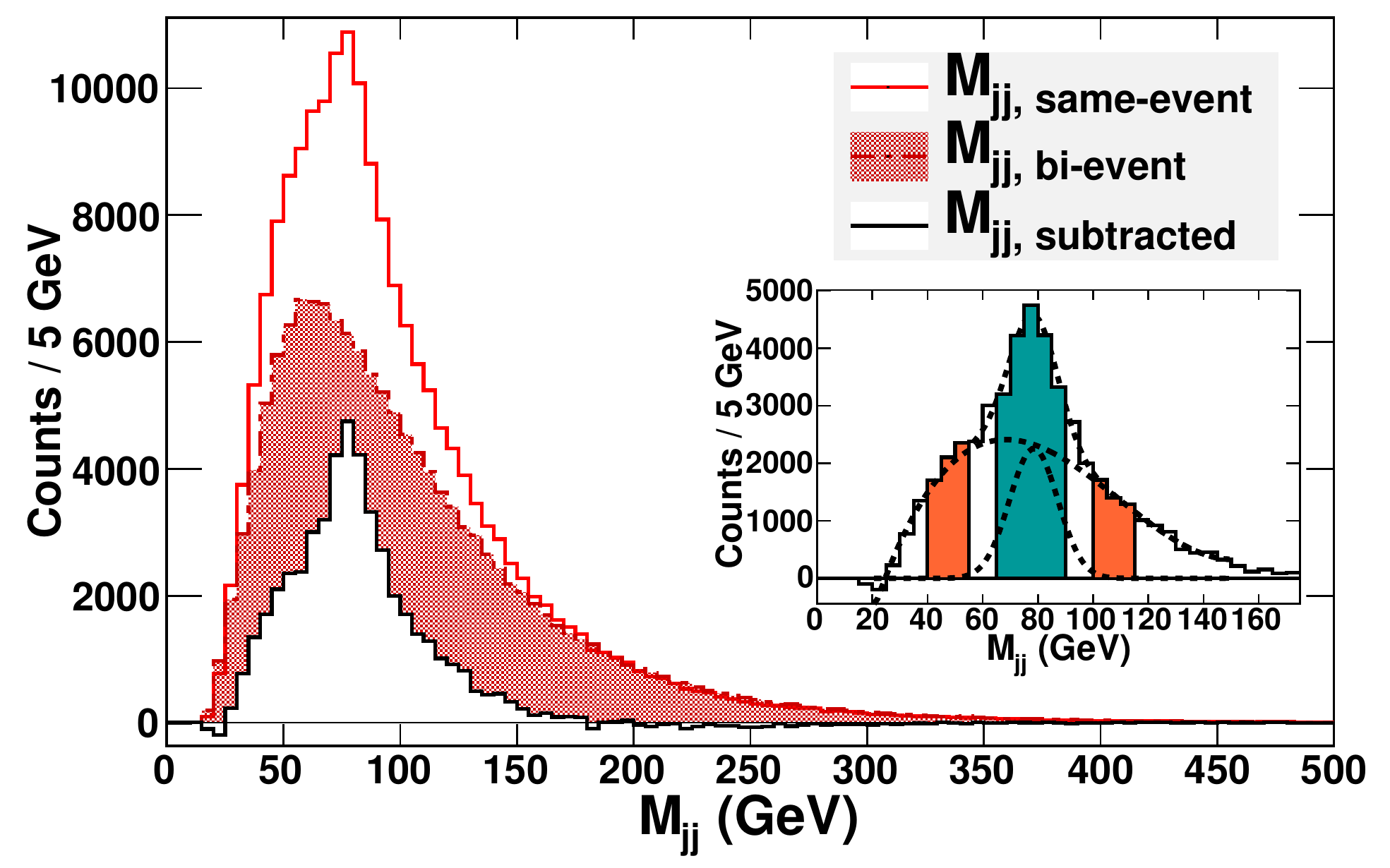}
\caption{The di-jet invariant mass distribution for our benchmark point. This figure demonstrates our BEST as well as how we find $W$ bosons in the events. The solid red(grey) histogram is constructed using same-event jets. The dot-dashed and filled red(grey) histogram is constructed using jets from different (or bi-) events and is normalized to the shape of the long tail in the same-event histogram. The same-event minus bi-event subtraction produces the black subtracted histogram. This subtracted histogram is plotted again in the inset plot. Also in the inset plot is shown the $W$ mass window (which is the cyan filled region between $65$ and $90~\gev$), the sideband windows (which are the orange filled regions between $40$ and $55~\gev$ as well as between $100$ and $115~\gev$), and the cubic plus Gaussian fit which describes the background and $W$ peak shapes (shown as short dashed lines).}
\label{figMjj}
\end{figure}

To help us find a clean $W$ peak with not too much background in the way, we also perform our BEST. To this end, we combine each jet considered with a jet from the previous event. Again, we sort these bi-event jet pairs into the $W$ mass window and the sideband window. We normalize the shape of the overall bi-event histogram by matching its tail with that of the same-event histogram and subtract. We are left with a $W$ peak surrounded by much less background. To subtract off the remaining background, we perform the sideband subtraction. We demonstrate this $W$ finding process, including our BEST and the sorting of jet pairs into the $W$ mass and sideband windows, in Fig.~\ref{figMjj}.

%%%%% Figure
%%%%%

\begin{figure} [t!]
\centering
\includegraphics[width=.70\textwidth]{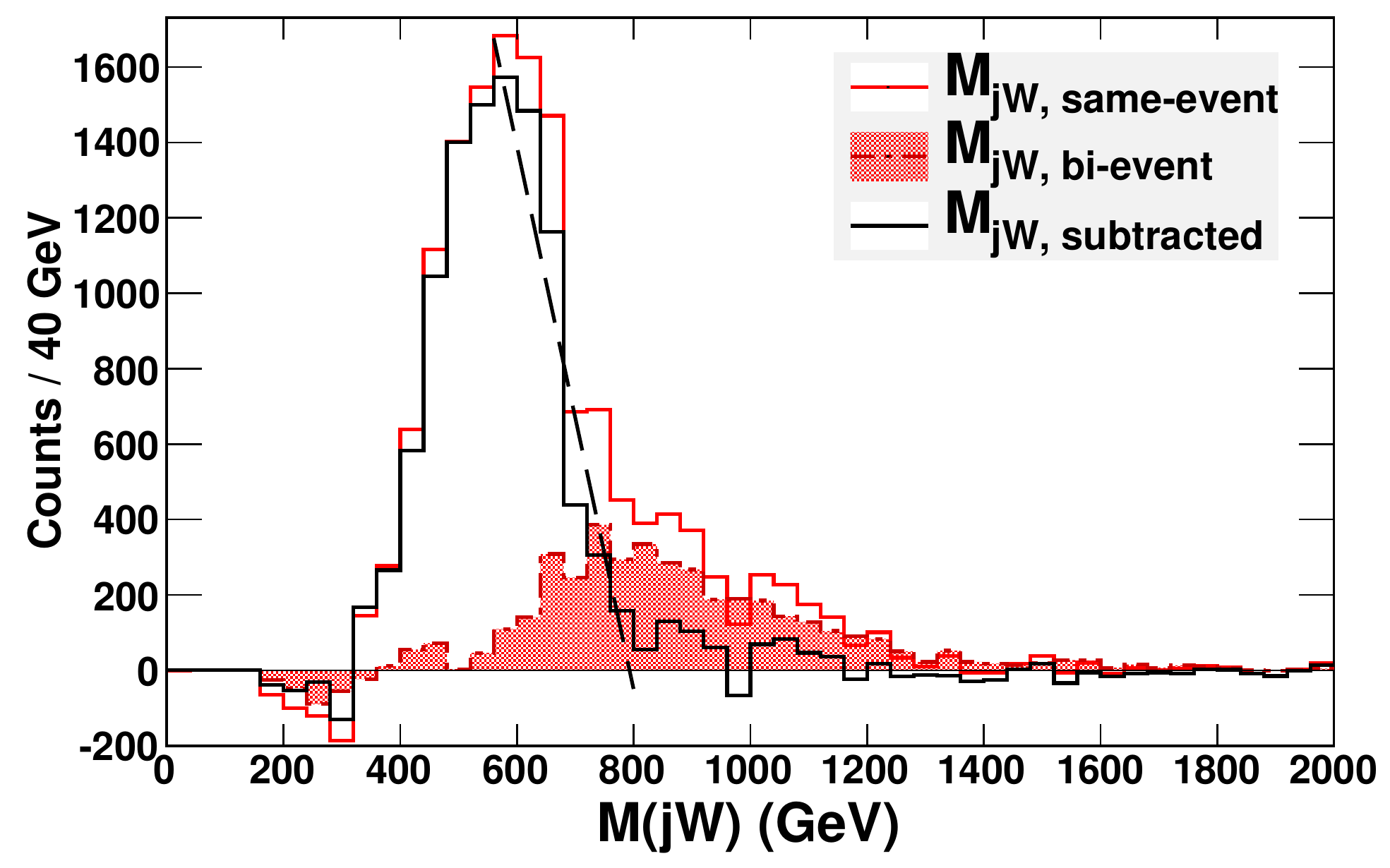}
\caption{The jet + $W$ invariant mass distribution for our benchmark point. This figure again demonstrates our BEST. The solid red(grey) histogram is constructed by combining each $W$ with each of the two leading jets from the same event. The dot-dashed and filled red(grey) bi-event histogram is constructed by combining each $W$ with each of the two leading jets from a different event and is normalized to the shape of the long tail in the same-event histogram. The same-event minus bi-event subtraction produces the black subtracted histogram. This subtracted histogram is then fitted with a straight line (shown as a dashed line in the figure) to find the endpoint of the distribution. The result for the endpoint from the fit is $793 \pm 2$(stat.)$~\pm 29$(syst.)$~\gev$.  This histogram and uncertainty are for an integrated luminosity of $1000~\invfb$.}
\label{figMjW}
\end{figure}

Once we implement this method of finding and reconstructing $W$ bosons, we begin to pair them up with leading jets. Once again, we perform our BEST to help us choose the correct leading jet (coming from the same cascade decay chain). Also, this second implementation of BEST significantly reduces the remaining SM background (which we tested by simulating SM backgrounds with \pythia), leaving a signal-to-background of around $10:1$.  So, we calculate the invariant mass $\mjW$ for our $W$ with leading jets from the same event, and again with leading jets from a different event. As before, we normalize the tail of the bi-event distribution to the shape of the tail of the same-event distribution and subtract. The final result is that the $\mjW$ distribution shows a nice endpoint which can be fit with a simple line. This endpoint becomes our $\mjW$ observable. A sample $\mjW$ distribution which shows our BEST is shown in Fig.~\ref{figMjW}.

%%%%% Figure
%%%%%

\subsection{The $\meff$ observable}

The effective mass observable~\cite{hinch1}, $\meff$, is a simple measure of the overall SUSY mass scale for many SUSY models. The signal is characterized by production of gluinos and squarks in the initial hard scattering events. The cascade decays of gluinos and squarks produce high transverse momentum jets and missing energy. The $\meff$ variable is defined as,
\begin{equation} %%%%% Equation
\meff = \pt_{\mathrm{jet},1} + \pt_{\mathrm{jet},2} + \pt_{\mathrm{jet},3} + \pt_{\mathrm{jet},4} + \met,
\label{eqMeffDefinition}
\end{equation} %%%%%
using the four highest transverse momentum ($\pt$) jets of the event. Because we select the leading four jets, we effectively get the jets which originate from gluino and squark decays.

We select events for the $\meff$ observable with the following cuts~\cite{hinch1}:
\begin{itemize}
\item Number of jets, $N_{\mathrm{jet}} \ge 4$ with jet $\pt \ge 50~\gev$ and $|\eta| \le 2.5$;
\item None of the above jets can be $b$ tagged;
\item Highest jet $\pt_{\mathrm{jet},1} \ge 100~\gev$;
\item No isolated $\mu$ leptons or electrons in the event;
\item Missing transverse energy, $\met \ge 200~\gev$ and $\met \ge 0.2\times\meff$;
\item Transverse sphericity, $S_T \le 0.2$.
\end{itemize}

\begin{figure} [t!]
\centering
\includegraphics[width=.70\textwidth]{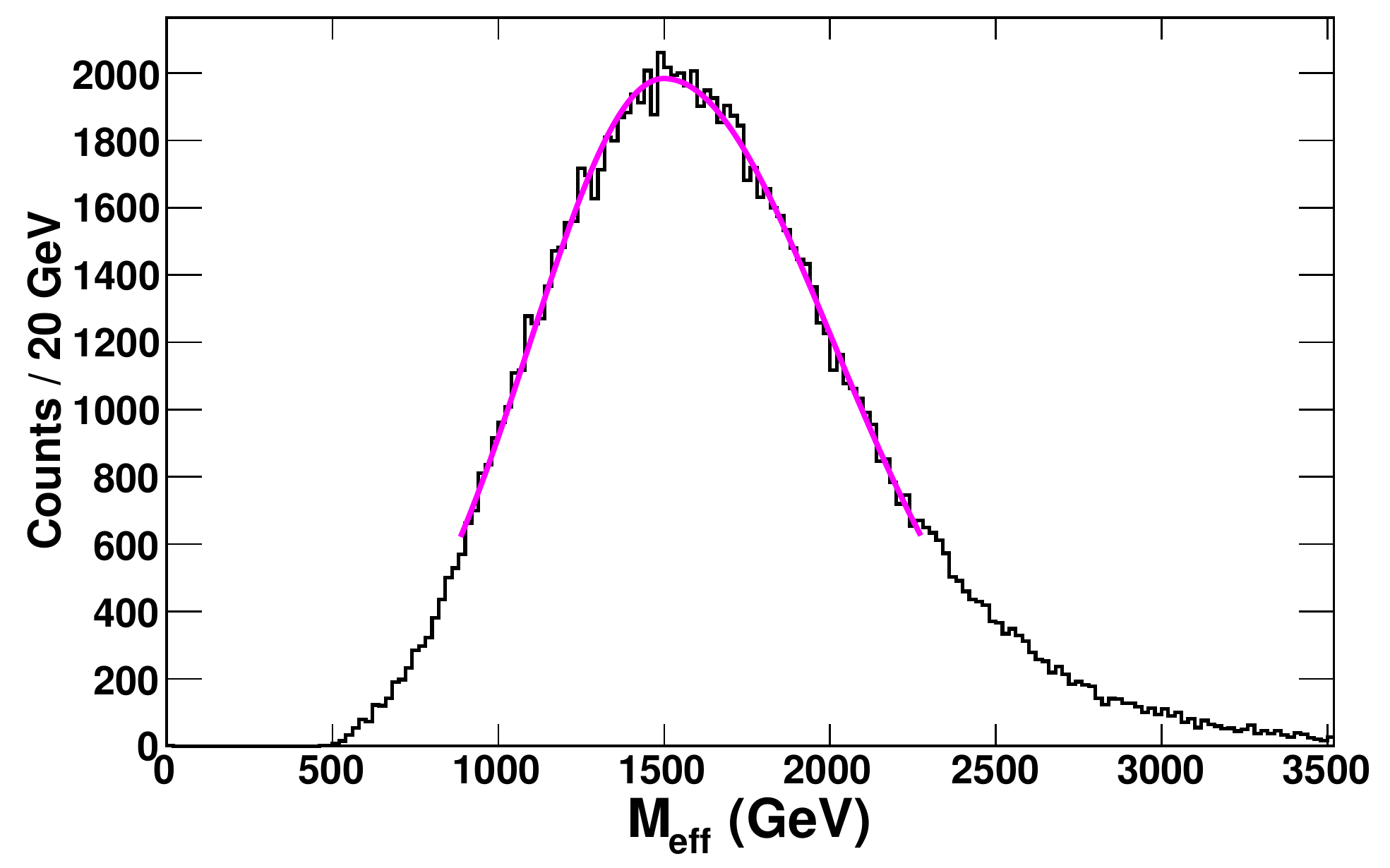}
\caption{The effective mass distribution for our benchmark point. The curve through the histogram is the result of fitting with the asymmetric Gaussian function given by Eq.~\ref{eqAsymGauss}. The result for the peak value from the fit is $1499 \pm 7$(stat.)$~\pm 45$(syst.)$~\gev$.  The histogram and uncertainty are for an integrated luminosity of $1000~\invfb$.}
\label{figMeff}
\end{figure}

Once events are selected in this way, the $\meff$ distribution is fit with an asymmetric Gaussian function:
\begin{equation} %%%%% Equation
N = \left\{ \begin{tabular}{l l}
        $C e^{-\frac{\meff - \meffpeak}{2\sigma_{\mathrm{low}}^2 } }$, & $\mathrm{if}\ \meff < \meffpeak$ \\
        $C e^{-\frac{\meff - \meffpeak}{2\sigma_{\mathrm{high}}^2 } }$, & $\mathrm{if}\ \meff \ge \meffpeak$ \\
      \end{tabular}\right. ,
\label{eqAsymGauss}
\end{equation} %%%%%
where $C$ is some constant scaling factor, $\meffpeak$ is the peak position, and $\sigma_{\mathrm{low}}$ and $\sigma_{\mathrm{high}}$ are the variances below and above the peak position, respectively. The most important result of the fit is the value of $\meffpeak$, which serves as our $\meff$ observable. A sample $\meff$ distribution is shown in Fig.~\ref{figMeff}.

%%%%% Figure
%%%%%

Another similar observable we define is $\meffbnoW$. This observable also measures the SUSY scale, but includes information from third generation squarks. It is defined as,
\begin{equation}  %%%%% Equation
\meffbnoW = \pt_{b\ \mathrm{jet},1} + \pt_{\mathrm{jet},2} + \pt_{\mathrm{jet},3} + \pt_{\mathrm{jet},4} + \met,
\label{eqMeffbDefinition}
\end{equation}  %%%%%
where in this case the leading jet (and only the leading jet) must be $b$ tagged. The selection cuts we use for this observable are identical to those of $\meff$ with the exception that no pair of jets in the entire event can have a invariant mass in the $W$ boson mass window. We define this mass window to be between $65$ and $90~\gev$. The $\meffbnoW$ distribution is fit in the same way as $\meff$. The plot of this distribution looks very similar in shape with the $\meff$ distribution shown in Fig.~\ref{figMeff}.

\subsection{Jet Plus $2\tau$}

The jet plus $2\tau$ signal originates from the following two decay chains:

\begin{equation} %%%%% Equation
  \squark \rightarrow q+\schitwozero(\schithreezero) \rightarrow q+\tau^{\mp}+\stauonepm \rightarrow q+\tau^{\mp}+\tau^{\pm}+\schionezero
\label{decayJetDitau}
\end{equation} %%%%%
The signal is characterized by a high $\pt$ jet and high $\met$ as well as a pair of oppositely charged $\tau$ leptons. 

We used the following cuts~\cite{LHCtwotau, LHCthreetau, LHCrelicdensity} to select events for this signal:
\begin{itemize}
\item Missing transverse energy, $\met \ge 180~\gev$;
\item At least two jets with jet $\pt \ge 200~\gev$ and $|\eta| \le 2.5$ (here we do not count $b$ tagged jets);
\item No $\mu$'s or electrons at all in the event;
\item At least two identified $\tau$ leptons~\cite{pgs} in the event with $\tau~\pt \ge 20~\gev$ and $|\eta| \le 2.5$;
\item The scalar sum, $\pt_{\mathrm{jet},1} + \pt_{\mathrm{jet},2} + \met \ge 600~\gev$;
\item There must be no $b$ tagged jet with $\pt$ larger than either of the two leading jets.
\end{itemize}
This signal can be utilized to make three independent observables, $\mtt$, $\mjtt$, and $\mjt$.  In this study, we treated these three observables as completely independent.  As such, we did not determine any correlation between their experimental uncertainties.

\subsubsection{$\mtt$}

\begin{figure} [t!]
\centering
\includegraphics[width=.70\textwidth]{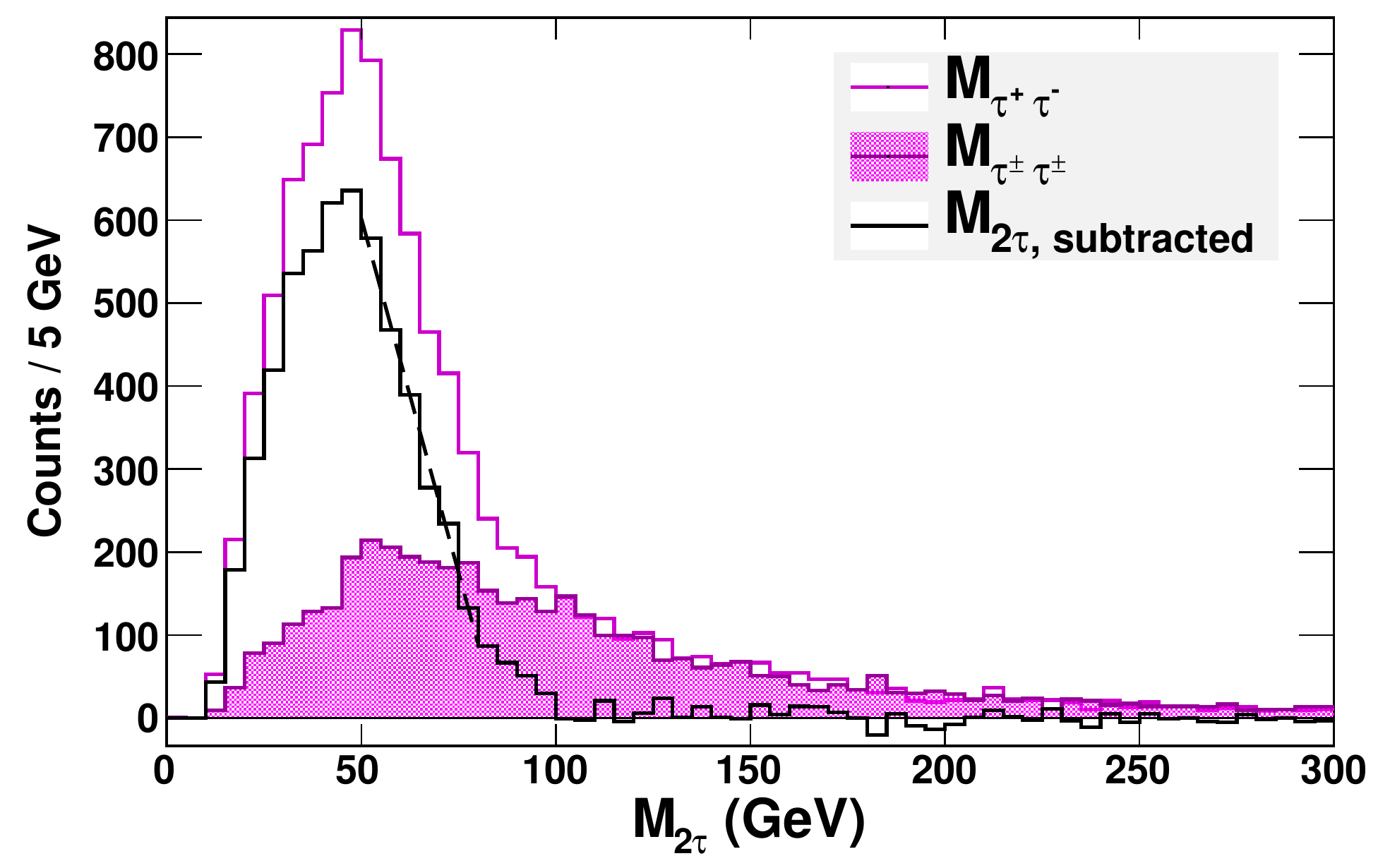}
\caption{The $2\tau$ invariant mass distribution for our benchmark point. The solid magenta(grey) histogram is composed of OS $\tau$ pairs, while the dot-dashed and filled magenta(grey) histogram is composed of LS $\tau$ pairs. The \osls subtraction produces the black subtracted histogram. This subtracted histogram is then fitted with a straight line (shown as a dashed line in the figure) to find the endpoint of the distribution. The result for the endpoint from the fit is $85.3 \pm 0.8$(stat.)$~\pm 3.8$(syst.)$~\gev$.  This histogram and uncertainty are for an integrated luminosity of $1000~\invfb$.}
\label{figMtt}
\end{figure}

To construct this observable we only need to combine $\tau$ pairs from each event. We sort the $\tau$ pairs into similarly charged or \textquotedblleft like-sign\textquotedblright\ (LS) pairs as well as oppositely charged or \textquotedblleft opposite-sign\textquotedblright\ (OS) pairs. The OS pairs contain $\tau$ pairs from our desired decay chain as well as random $\tau$ pairs, whereas the LS pairs contain only random $\tau$ pairs. Thus, we perform an \osls subtraction to make the $2\tau$ invariant mass, $\mtt$. This distribution shows a nice endpoint which can be determined by fitting with a simple line. A sample distribution showing the \osls subtraction is shown in Fig.~\ref{figMtt}. Due to the inability to fully reconstruct each $\tau$ (because of neutrinos which are missing energy), a small shoulder appears in the figure beyond the endpoint. However, this shoulder is not in the way of finding the endpoint for this case.  This endpoint serves as our $\mtt$ observable.

%%%%% Figure
%%%%%

\subsubsection{$\mjtt$}

\begin{figure} [t!]
\centering
\includegraphics[width=.70\textwidth]{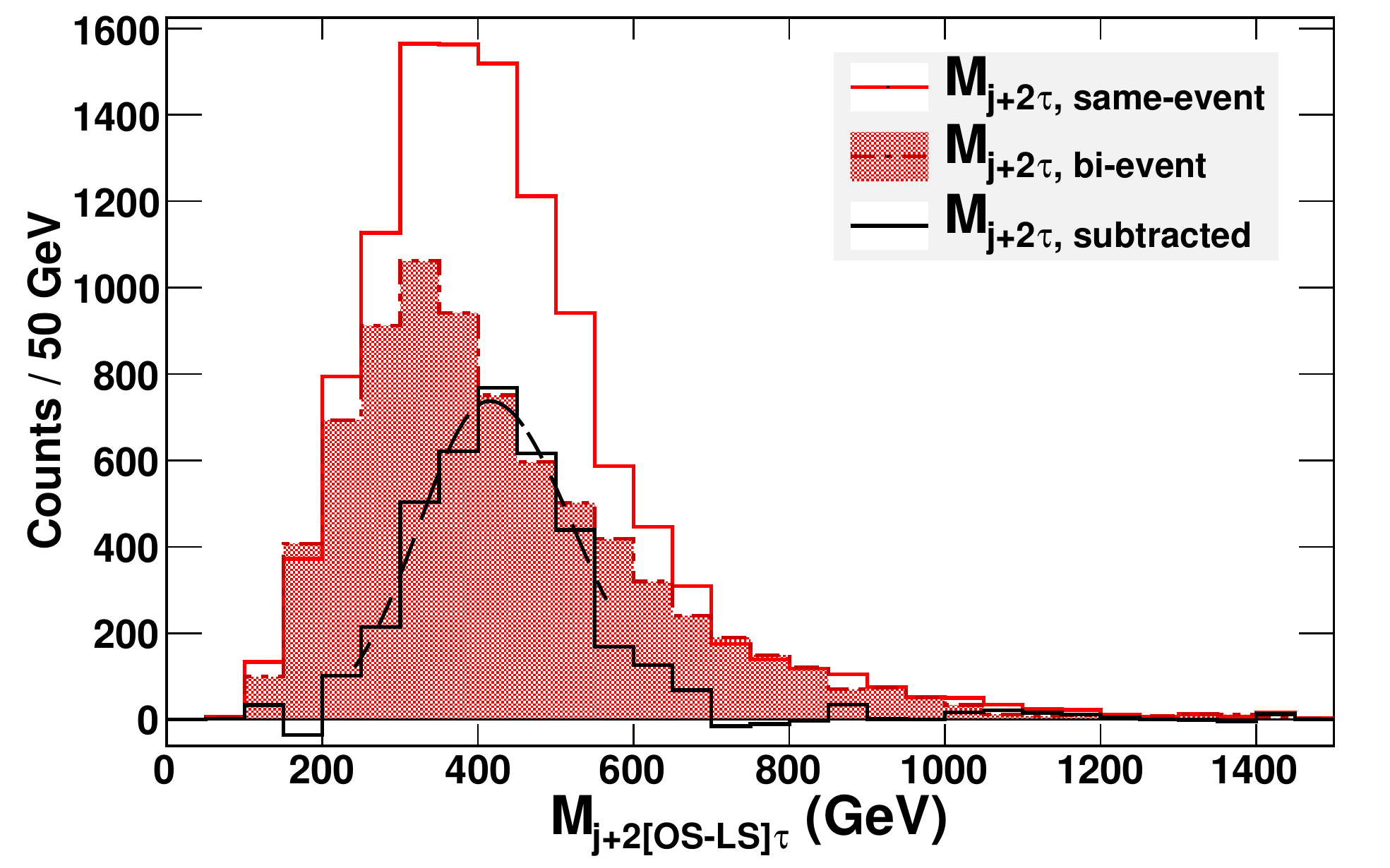}
\caption{The jet + $2\tau$ invariant mass distribution for our benchmark point. The solid red(grey) histogram is constructed by combining each \osls $\tau$ pair with each of the two leading jets from the same event. The dot-dashed and filled red(grey) bi-event histogram is constructed by combining each \osls $\tau$ pair with each of the two leading jets from a different event and is normalized to the shape of the long tail in the same-event histogram. The same-event minus bi-event subtraction produces the black subtracted histogram. This subtracted histogram is then fitted with a Gaussian function (shown as a dashed curve in the figure) to find the peak of the distribution. The result for the peak value from the fit is $415 \pm 8$(stat.)$~\pm 40$(syst.)$~\gev$.  This histogram and uncertainty are for an integrated luminosity of $1000~\invfb$.}
\label{figMjtt}
\end{figure}

As for $\mtt$, we consider all $\tau$ pairs from each event, sorting them into OS and LS pairs. We combine each pair with each of the leading jets to fill the same-event jet plus $2\tau$ invariant mass distribution, $\mjttsame$. Additionally, each $\tau$ pair is combined with each of the two leading jets from a different event to fill the bi-event distribution, $\mjttbi$. The \osls subtraction is performed first, followed by our BEST. As before, the bi-event histogram is normalized to the shape of the tail in the same-event histogram prior to the subtraction. The result of this process is the $\mjtt$ distribution. Unlike our $\mtt$ observable, we do not see a sharp endpoint. Due to the effects of missing energy from each $\tau$ (when the neutrino escapes the detector) as well as combining those $\tau$'s with a jet, the endpoint gets washed out. Instead, we choose the peak position for our $\mjtt$ observable. This peak is found by fitting with either a regular Gaussian or asymmetric Gaussian (seen in Eq.~\ref{eqAsymGauss}), depending on the shape of the distribution. The fitted peak value serves as our $\mjtt$ observable. A sample distribution showing our BEST is shown in Fig.~\ref{figMjtt}.

%%%%% Figure
%%%%%

\subsubsection{$\mjt$}

To construct this observable we consider $\tau$ pairs in the event and combine one of the $\tau$'s with the corresponding leading jets. There is an ambiguity which arises due to the choice of which $\tau$ to use in the observable. The way we deal with this ambiguity is described below.

Similarly as before we collect all possible $\tau$ pairs in the event, sorting them into OS and LS pairs. For each leading jet, the invariant mass, $\mjt$, is determined for each tau. These two values are compared, and stored in two histograms. The histograms are labeled as $\mjtFirst$, which contains the larger of the two $\mjt$ values for each jet, and $\mjtSecond$, which contains the smaller. A similar procedure is performed for the two leading jets from a different event to form bi-event distributions. The \osls subtraction is performed, followed by our BEST. This leaves us with two resulting final distributions of $\mjtFirst$ and $\mjtSecond$.

\begin{figure} [t!]
\centering
\includegraphics[width=.70\textwidth]{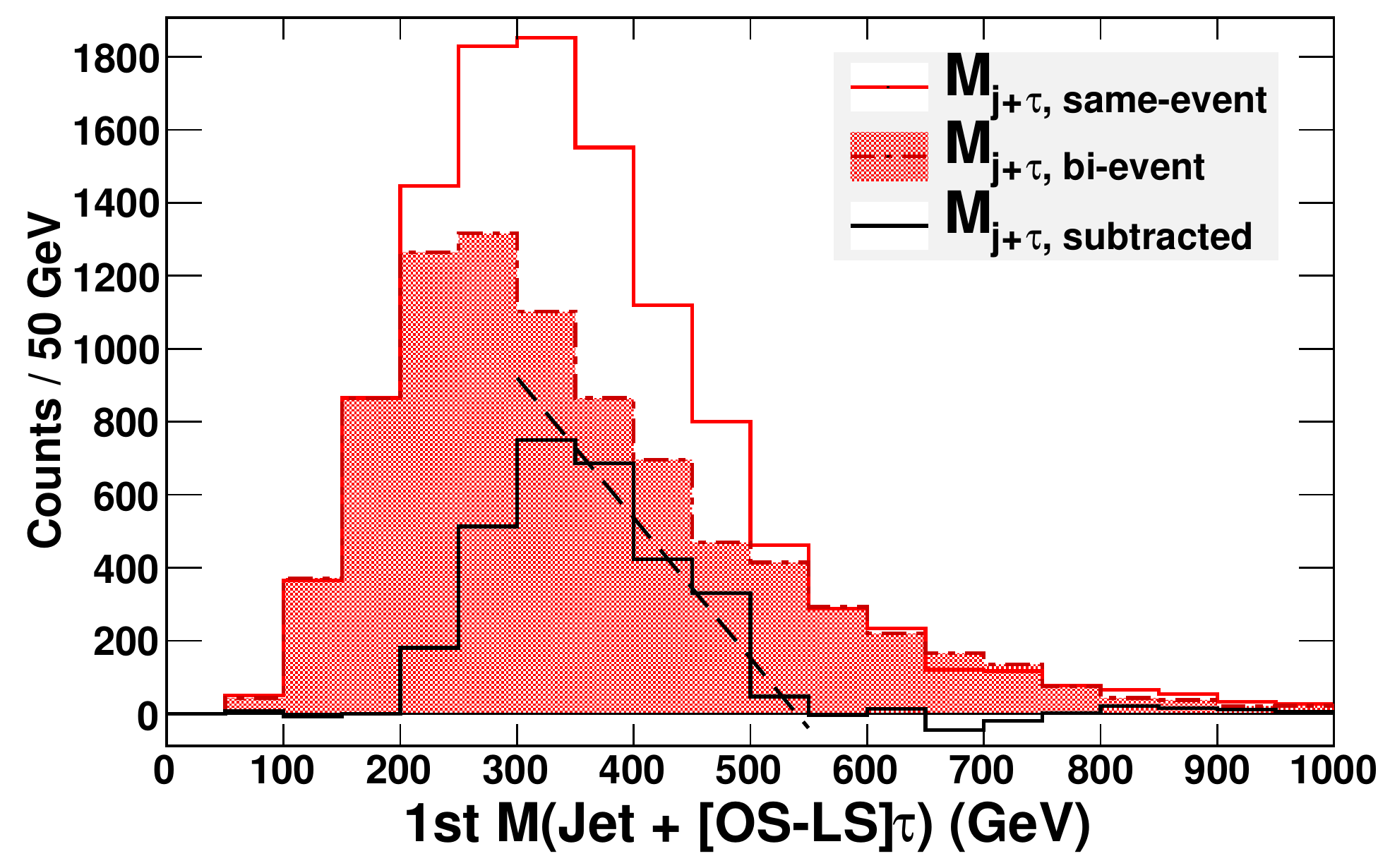}
\caption{The jet + $\tau$ invariant mass distribution for our benchmark point. The solid red(grey) histogram is constructed by combining each \osls $\tau$ with each of the two leading jets from the same event, choosing the $\tau$ which gives the larger value of $\mjt$. The dot-dashed and filled red(grey) bi-event histogram is constructed the same way with each of the two leading jets from a different event and is normalized to the shape of the long tail in the same-event histogram. The same-event minus bi-event subtraction produces the black subtracted histogram. This subtracted histogram is then fitted with a linear function (shown as a dashed line in the figure) to find the endpoint of the distribution. The result for the endpoint from the fit is $540 \pm 2$(stat.)$~\pm 34$(syst.)$~\gev$.  This histogram and uncertainty are for an integrated luminosity of $1000~\invfb$.}
\label{figMjt}
\end{figure}

For this region of parameter space, there is a systematic way to choose between these histograms to find an endpoint which is close in agreement with the theoretical prediction. By default, the $\mjtFirst$ histogram is chosen and fit with a line to find the endpoint of the distribution. However, if the behavior of the endpoint region does not seem linear, we choose $\mjtSecond$. We also make this choice if the BEST does not appear to work. The BEST can fail if there are not enough signal events compared to background events. If the BEST fails like this for $\mjtFirst$, it tells us we are not picking up the decay chain we want to look at, so we choose $\mjtSecond$ instead. A sample distribution for our benchmark point is shown in Fig.~\ref{figMjt}. Since we have less missing energy here (only one $\tau$), we can see the endpoint clearly for this observable.  This endpoint is our $\mjt$ observable.

%%%%% Figure
%%%%%

\section{Determining Model Parameters and Relic Density}

With the six observables in section 3, we determine the model parameters. We vary each model parameter independently about our benchmark point to find how each observable behaves as a function of the model parameters. We vary one parameter at a time, performing at least three additional simulations for each parameter.  If an observable is found to be particularly non-linear in a certain region, we simulate additional points in that region.  For instance, to vary the $\mhalf$ parameter alone, we held all the other parameters fixed and performed simulations for $\mhalf = \{480, 490, 520, 540\}~\gev$ in addition to our base point. To get a five dimensional cross in parameter space, we did the same for all the five parameters, requiring around 30 simulations in total. 

We simulate the LHC experiment for each point in our cross, and determine all the observables along with their uncertainties. A sample set of observable results and uncertainties for our benchmark point is shown in Table~\ref{tabObservables}. For each observable, we plot the value of the observable as a function of each model parameter. Fits through these plots determine the \textquotedblleft functional form\textquotedblright\ for our observables in a similar fashion as in Ref.~\cite{LHCrelicdensity}. In this region of nuSUGRA parameter space, the observables behave in a fortunate manner. Some observables are only functions of some of the model parameters, while being constant with respect to variations in other model parameters. The functional forms are as follows:
\begin{itemize}
\item $\meffpeak = f_1(\mhalf)$;
\item $\meffbnoWpeak = f_2(\mhalf)$;
\item $\mjWend = f_3(\mhalf, m_H)$;
\item $\mjttpeak = f_4(\mhalf, m_H, \mzero)$;
\item $\mttend = f_5(\mhalf, m_H, \mzero, \azero)$;
\item $\mjtend = f_6(\mhalf, m_H, \mzero, \azero, \tanb)$.
\end{itemize}

\begin{table}
\caption{Results from the fits of kinematical observables found at our benchmark point, along with its statistical uncertainty for luminosities of $1000~\invfb$ and $100~\invfb$, and its systematic uncertainty~\cite{uncertainties}. All values have units of $\gev$.}
\label{tabObservables}
\begin{center}
\begin{tabular}{c | c c c c}
\hline \hline
Observable & Value & $1000~\invfb$ Stat. & $100~\invfb$ Stat. & Systematic \\ \hline
$\meffpeak$ & 1499 & $\pm 7$ & $\pm 21$ & $\pm 45$\\
$\meffbnoWpeak$ & 1443 & $\pm 43$ & $\pm 107$ & $\pm 43$\\ 
$\mjWend$ & 793 & $\pm 2$ & $\pm 5$ & $\pm 29$\\
$\mjttpeak$ & 415 & $\pm 8$ & $\pm 26$ & $\pm 40$ \\ 
$\mttend$ & 85.3 & $\pm 0.8$ & $\pm 2.8$ & $\pm 3.8$ \\
$\mjtend$ & 540 & $\pm 2$ & $\pm 6$ & $\pm 34$ \\ \hline \hline
\end{tabular}
\end{center}
\end{table}

These functional forms make a lot of sense. The sensitivity of each observable on a model parameter depends on the relative change that varying that parameter will affect the observable. For instance, the $\meff$ observables have peak values that are quite large ($\simeq 1500~\gev$), so small changes in the squark masses (caused by varying $\mzero$, $\azero$, or $\tanb$) are undetected. Thus, in this region of parameter space, the $\meff$ observables are only a function of $\mhalf$. A similar effect happens for $\mjW$ as well. However, $\mjW$ also depends on neutralino and chargino masses which are strongly affected by $\mu(m_H)$. Both squark masses and neutralino masses show a slight $\mzero$ dependence. Since, for $\mjtt$, we are looking at the relatively small peak position ($\simeq 400~\gev$), we pick up this $\mzero$ sensitivity. Lastly, the $\mtt$ and $\mjt$ feel the involvement of the stau particle, showing dependencies on $\mzero$, $\azero$, and $\tanb$.

This behavior luckily allows us to solve for one parameter at a time. If we wanted to solve for all parameters at once, perhaps performing a least squares fit, that would require many additional simulations. We would have to simulate a grid in parameter space rather than just a cross. Thus, the advantage of solving for one parameter at a time whenever it is possible is to save on computing time. Perhaps in future studies which accompany actual LHC data, the least squares fit method would be preferable. The least squares method to find the model parameters would find correlations between the uncertainties. In addition, the least squares method would be potentially more precise. However, for our feasibility study, finding the parameters one at a time is sufficient.

The method to solve for one parameter at a time is as follows. The result of each model parameter is used as input for the next parameter to be solved for. In this manner, solving for each parameter is as simple as solving for one unknown from one equation. However, the uncertainty in each solved parameter would then influence the next one to be solved for. To estimate this effect, we use the uncertainty in each parameter as an additional source of uncertainty for the next observable. All such uncertainties are estimated using simple Monte Carlo programs.

To illustrate this whole process, we describe the first few logical steps in how we determine the model parameters: $\meffpeak$ and $\meffbnoWpeak$ are only functions of $\mhalf$, so we use each to solve for $\mhalf$ separately, then combine the measurements. This measurement combination reduces the uncertainty slightly (around a 7\% uncertainty reduction) as compared to using $\meffpeak$ alone. Next, we propagate the uncertainty in $\mhalf$ to an additional uncertainty in $\mjWend$ by using the $\mjWend$ versus $\mhalf$ functional form. This uncertainty is added in quadrature to the measurement uncertainty in $\mjWend$. With the uncertainty in $\mjWend$ estimated this way, we use the $\mjWend$ versus $m_H$ functional form to solve for $m_H$. Then the uncertainties in $\mhalf$ and $m_H$ are propagated as additional uncertainties in $\mjttpeak$ while solving for $\mzero$. The process continues like this all the way down the above list of functional forms.

Once we have finally determined all the model parameters, we use \darksusy~\cite{darksusy} to calculate the dark matter relic density of the universe today, $\DMrelic$. We also estimate the uncertainty in the dark matter relic density due to the uncertainties in the measured model parameters. Our results are shown in Table~\ref{tabResults}. We find that the model parameters $\mzero$, $\mhalf$, $m_H$ and $\tanb$ can be determined a good accuracy: The statistical uncertainties are $\lesssim 15\%$ for $100~\invfb$ luminosity, with the systematic uncertainties nearly the same. The relative uncertainty in the parameter $m_0$ is somewhat larger than the others due to the fact that we determine it with the $\mjttpeak$ observable. The peak value of $\mjtt$ is a less accurate measure of the SUSY masses in the decay chain (Equation~\ref{decayJetDitau}) than the endpoint would be, were it possible to use the endpoint. 

We can determine the accuracy of $\mu$ from these parameters and we find that $\mu$ can be determined with accuracies of around 15\% and 8\% for luminosities of $100~\invfb$ and $1000~\invfb$, respectively. The uncertainty of $\mu$ is influenced not only by the uncertainty in $m_H$, but by $\mzero$ and other model parameters as well, as surmised from Equation~\ref{eqMu}. Even though that equation is for low and intermediate $\tanb$, we see in Table~\ref{tabResults} a similar behavior that the uncertainty of $\mu$ is dominated by the uncertainty in $\mzero$ which is large compared to other model parameters.

\begin{figure} [t!]
\centering
\includegraphics[width=.70\textwidth]{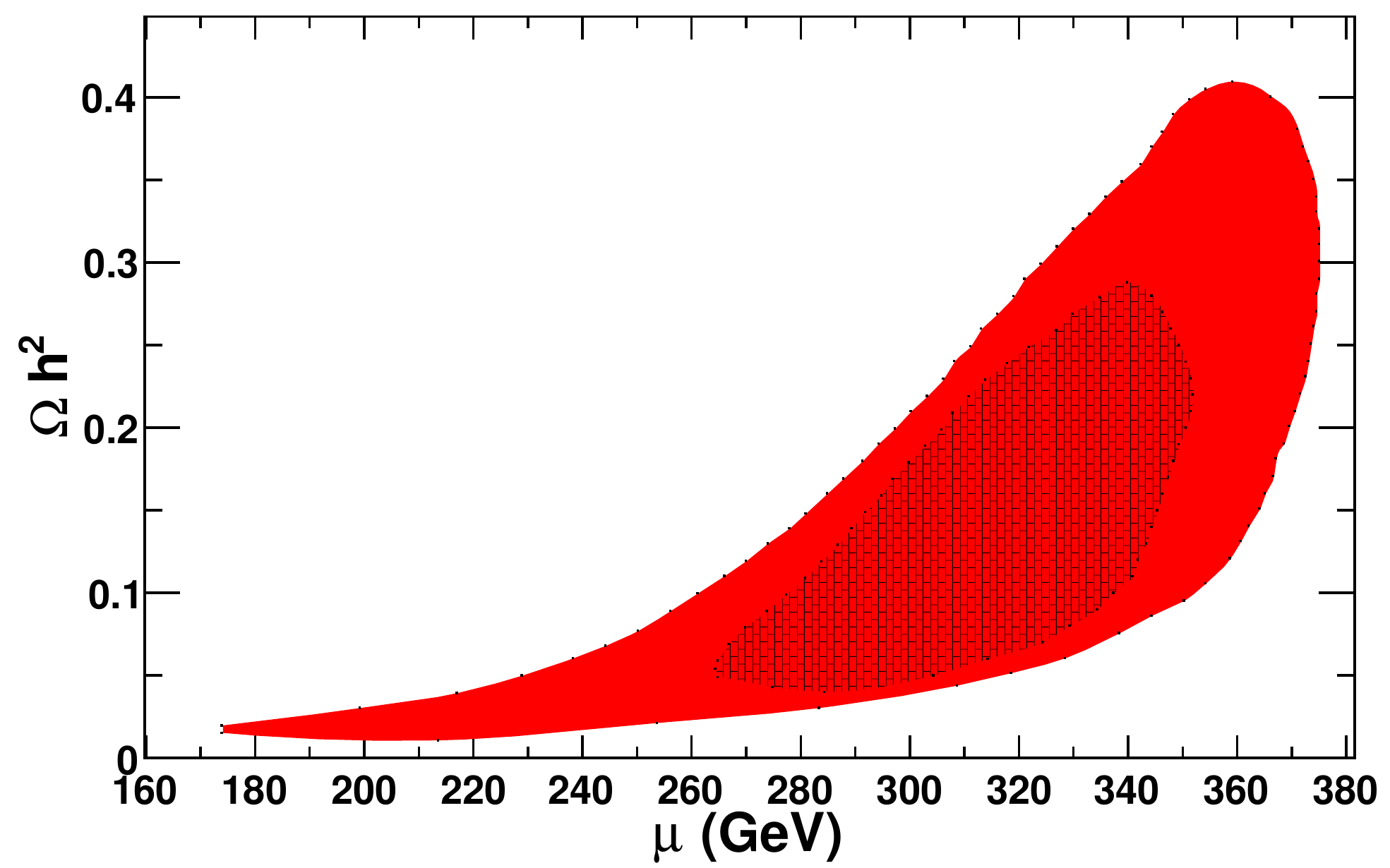}
\caption{Estimates of the statistical $1\sigma$ uncertainties in the $\DMrelic$ versus $\mu$ plane. The solid red (brick textured) region is for a luminosity of $100~\invfb$ ($1000~\invfb$).}
\label{figEllipse}
\end{figure}

Since the dark matter content is sensitive to the value of $\mu$, in Fig.~\ref{figEllipse}, we plot one $\sigma$ contours of the dark matter content as a function of $\mu$ for luminosities of $100~\invfb$ (red shaded region) and $1000~\invfb$ (brick shaded region). The determination of dark matter content is of couse much better with $1000~\invfb$, but even with $100~\invfb$ the measurement accuracy is quite encouraging.

\begin{table}
\caption{Results of the nuSUGRA model parameters and relic density of dark matter in the universe for integrated luminosities of $1000~\invfb$ and $100~\invfb$. The systematic uncertainties are also estimated here~\cite{uncertainties}. Note that the uncertainties for an integrated luminosity of $100~\invfb$ were estimated by simply scaling down the distributions before performing fits for the analysis.}
 \label{tabResults}
\begin{center}
\begin{tabular}{c | c c c c c | c c}
\hline \hline
$\lum~(\invfb)$ & $\mhalf~(\gev)$ & $m_H~(\gev)$ & $\mzero~(\gev)$ &
$\azero~(\gev)$ & $\tanb$ & $\mu~(\gev)$ & $\DMrelic$ \\ \hline
$1000$ & $500 \pm 3$ & $727 \pm 10$ & $366 \pm 26$ & 
$3 \pm 34$ & $39.5 \pm 3.8$ & $321 \pm 25$ & $0.094^{+0.107}_{-0.038}$ \\ 
$100$ & $500 \pm 9$ & $727 \pm 13$ & $367 \pm 57$ & 
$0 \pm 73$ & $39.5 \pm 4.6$ & $331 \pm 48$ & $0.088^{+0.168}_{-0.072}$ \\ 
Syst. & $\pm 10$ & $\pm 15$ & $\pm 56$ & 
$\pm 66$ & $\pm 4.5$ & $\pm 48$ & $ ^{+0.175}_{-0.072}$ \\ \hline \hline
\end{tabular}
\end{center}
\end{table}

%%%%% Figure
%%%%%

\section{Conclusions and Discussion}

In this paper we have shown that the LHC has an ability to investigate the origin of dark matter by establishing SUSY models. At the LHC, the colored SUSY particles, the squarks and gluinos, will be produced profusely. The squarks and gluinos will then go through cascade decays into final states involving the SM particles and missing energy. The challenge is to reconstruct masses or model parameters by forming observables using these final states. If we can reconstruct the model we will be able to calculate the dark matter content and check whether the established model explains the cosmology correctly. In this way, establishing the MSSM could be very hard. We may not have enough observables to measure all the parameters of the MSSM. Also not all the particles of the model would show up in the cascade decays with sufficient branching ratios. We therefore started with a simpler model where the gaugino masses are unified, and all scalar masses except that of the Higgs bosons are also unified at the GUT scale. This type of model is quite realistic and very popular since the explanation of dark matter becomes easy. The explanation can occur either via a large Higgsino component in the neutralino or via a resonance with the heavy or pseudoscalar Higgs. Here we considered the scenario with the larger Higgsino component. This scenario also gives rise to a larger direct detection cross-section and therefore can be detected in the upcoming direct detection experimental results.

Since SUSY production occurs in pairs of colored SUSY particles, each event has two decay chains. The particle from one decay chain will create background for any measurement of mass involving the other decay chain. In this paper, we first established techniques to remove this kind of background by combining particles from different events. We then created observables to establish the model. In previous works, when we tried to establish the mSUGRA model, we found that the observables mostly involved leptons, jets, Higgs bosons and $Z$ bosons. However, for this nonuniversal Higgs model, nuSUGRA, we found that the final states involved $W$ bosons arising from chargino decays in addition to jets, and $\tau$ leptons in the final states. We constructed observables, e.g., $Wj$, $j\tau\tau$, $\tau\tau$ etc. and use them to determine masses and the model parameters. We showed that the model parameters can be determined with good accuracy (e.g. $m_0$, $m_{1/2}$, $m_H$, and $\tanb$ have statistical and systematic uncertainties of $\lesssim 15\%$ for a combined uncertainty of $\lesssim 20\%$) at a luminosity of $100~\invfb$. The parameter $\mu$ can be determined with an accuracy of 21\% for the same luminosity. Finally, we showed that the dark matter content also can be determined in agreement with the WMAP experiment.

\section*{Acknowledgments}
This work is supported in part by the DOE grant DE-FG02-95ER40917 and by the World Class University (WCU) project through the National Research Foundation (NRF) of Korea funded by the Ministry of Education, Science \& Technology (grant No. R32-2008-000-20001-0). We would like to thank A. Gurrola and Y. Santoso for useful discussions.


\begin{thebibliography}{99}

\bibitem{WMAP}
WMAP Collaboration, D.N. Spergel {\it et al.}, Astrophys. J. Suppl. {\bf 148} (2003) 175.

\bibitem{neuDM}H. Goldberg, Phys. Rev. Lett. \textbf{50} (1983) 1419;
 J. Ellis, J. Hagelin, D. Nanopoulos, K. Olive, and M. Srednicki,
 Nucl. Phys. B \textbf{238} (1984) 453.

\bibitem{sps} B. C. Allanach \etal, [arXiv:hep-ph/0403133];
P. Bechtle, K. Desch, W. Porod, and P. Wienemann, Eur. Phys. J. C \textbf{46}, (2006) 533;
R. Lafaye, T. Plehn, M. Rauch and D. Zerwas, Eur. Phys. J. C \textbf{54}, (2008) 617;
J. L. Kneur and N. Sahoury, Phys. Rev. D \textbf{79}, (2009) 075010;
C. Adam \etal, [arXiv:hep-ph/1007.2190]. \\


\bibitem{sugra1}D. Z. Freedman, P. Van Niewenhuisen, and S. Ferrara, Phys. Rev. D \textbf{13} (1976) 3214; S. Deser and B. Zumino, Phys. Lett. B \textbf{65} (1976) 369;
A.H. Chamseddine, R. Arnowitt, and P. Nath,  Phys. Rev. Lett. \textbf{49} (1982) 970;
R. Barbieri, S. Ferrara, and C.A. Savoy, Phys. Lett. B \textbf{119} (1982) 343; L. Hall, J. Lykken, and S. Weinberg, Phys. Rev. D \textbf{27} (1983) 2359; P. Nath, R. Arnowitt, and A.H. Chamseddine, Nucl. Phys. B \textbf{227} (1983) 121; For a review, see P. Nilles, Phys. Rep. \textbf{110} (1984) 1.

\bibitem{bsgamma}
M. Alam \etal, Phys. Rev. Lett. \textbf{74} (1995) {2885}.

\bibitem{LHCtwotau}
R. Arnowitt \etal,
%``Detection of SUSY in the stau-neutralino coannihilation region at the
 %LHC,''
 Phys. Lett. B \textbf{639} (2006) 46.
% [arXiv:hep-ph/0603128].
 %%CITATION = PHLTA,B639,46;%%Phys. Lett. B {\bf 639} (2006) 46.

\bibitem{LHCthreetau}
R. Arnowitt \etal,
%``Indirect measurements of the stau - neutralino 1(0) mass difference and
 %mSUGRA in the co-annihilation region of mSUGRA models at the LHC,''
 Phys. Lett. B \textbf{649} (2007) 73.
 %%CITATION = PHLTA,B649,73;%%

\bibitem{LHCrelicdensity}
R. Arnowitt \etal,
%``Determining the Dark Matter Relic Density in the mSUGRA Stau-Neutralino Co-Annihilation Region at the LHC,''
Phys. Rev. Lett. \textbf{100} (2008) 231802.
% [arXiv:hep-ph/0802.2968v1]. Accepted for publication in PRL.

\bibitem{sscprd}
B. Dutta \etal,
%``Supersymmetry signals of supercritical string cosmology at the LHC''
 Phys. Rev. D \textbf{79} (2009) 055002.
 %%CITATION = PHLTA,D79,055002;%%
 
\bibitem{nath1}P. Nath and R. Arnowitt, Phys. Rev. \textbf{D56} (1997) 2820.

\bibitem{nuhm}
See for example:\\ 
H. Baer, A. Mustafayev, S. Profumo, A. Belyaev, and X. Tata,
 %``Neutralino cold dark matter in a one parameter extension of the minimal
  %supergravity model,''
  Phys.\ Rev.\  D {\bf 71} (2005) 095008 and JHEP {\bf 0507} (2005) 065;
%  [arXiv:hep-ph/0412059].
 %%CITATION = PHRVA,D71,095008;%%
  J. R. Ellis, K. A. Olive, and Y. Santoso,
  %``The MSSM Parameter Space with Non-Universal Higgs Masses,''
  Phys.\ Lett.\  B {\bf 539} (2002) 107 and
  J. High Energy Phys. \textbf{0810} (2008) 005;
 % [arXiv:hep-ph/0204192].
  %%CITATION = PHLTA,B539,107;%%
 J. R. Ellis, T. Falk, K. A. Olive, and Y. Santoso,
  %``Exploration of the MSSM with Non-Universal Higgs Masses,''
  Nucl.\ Phys.\  B {\bf 652} (2003) 259;
  J. R. Ellis, K. A. Olive, and P. Sandick,
  New J. Phys. \textbf{11} (2009) 105015;
  J. R. Ellis, S. F. King, and J. P. Roberts,
  J. High Energy Phys. \textbf{0804} (2008) 099;
 A. Bottino, F. Donato, N. Fornengo, and S. Scopel,
  %``Compatibility of the new DAMA/NaI data on an annual modulation effect  in
  %WIMP direct search with a relic neutralino in supergravity schemes,''
  Phys.\ Rev.\  D {\bf 59} (1999) 095004 and Phys.\ Rev.\  D {\bf 63} (2001) 125003;
 E. Accomando, R. Arnowitt, B. Dutta, and Y. Santoso,
  %``Neutralino proton cross-sections in supergravity models,''
  Nucl.\ Phys.\  B {\bf 585} (2000) 124;
 S. Bhattacharya, U. Chattopadhyay, D. Choudhury, D. Das, B. Mukhopadhyaya,
Phys. Rev. D \textbf{81} (2010) 075009;
L. Roszkowski \etal,
hep-ph/0903.1279;
U. Chattopadhyay and D. Das,
Phys. Rev. D \textbf{79} (2009) 035007;
A. De Roeck \etal,
Eur. Phys. J. C \textbf{49} (2007) 1041;
D. G. Cerdeno and C. Munoz,
J. High Energy Phys. \textbf{0410} (2004) 015.



 

\bibitem{darkrv}
J. Ellis, K. Olive, Y. Santoso, and V. Spanos,
Phys. Lett. B \textbf{565} (2003) {176}; 
R. Arnowitt, B. Dutta, and B. Hu,
hep-ph/0310103; 
H. Baer \etal, 
{J. High Energy Phys.} \textbf{0306} (2003) {054}; 
A. B. Lahanas and D.V. Nanopoulos, 
Phys. Lett. B \textbf{568} (2003) {55}; 
U. Chattopadhyay, A. Corsetti, and P. Nath, 
Phys. Rev. D \textbf{68} (2003) {035005}; 
E. Baltz and P. Gondolo, 
J. High Energy Phys. \textbf{0410} (2004) 052; 
A. Djouadi, M. Drees, and J. L. Kneur,
  %``Updated constraints on the minimal supergravity model,''
  J. High Energy Phys. \textbf{0603} (2006) 033;
%  [arXiv:hep-ph/0602001].
  %%CITATION = J. High Energy Phys.A,0603,033;%%
%J. L. Feng, K. T. Matchev, and F. Wilczek,
  %``Neutralino dark matter in focus point supersymmetry,''
%  Phys. Lett. B \textbf{482} (2000) 388;
  %[arXiv:hep-ph/0004043].
  %%CITATION = PHLTA,B482,388;%%
 G. Belanger, S. Kraml, and A. Pukhov,
  %``Comparison of SUSY spectrum calculations and impact on the relic  density
  %constraints from WMAP,''
  Phys. Rev. D \textbf{72} (2005) 015003.
 % [arXiv:hep-ph/0502079].
  %%CITATION = PHRVA,D72,015003;%%

\bibitem{stauco}J. Ellis, T. Falk, and K. Olive, Phys. Lett. B \textbf{444} (1998)
367; J. Ellis, T. Falk, K. Olive, and
M. Srednicki, Astropart. Phys. \textbf{13} (2000) 181; M.E. G«omez, G. Lazarides,
and C. Pallis, Phys.
Rev. D \textbf{61} (2000) 123512 and Phys. Lett. B \textbf{487} (2000) 313; A. B. Lahanas, D.
V. Nanopoulos,
and V. C. Spanos, Phys. Rev. D \textbf{62} (2000) 023515; R. Arnowitt, B. Dutta, and Y.
Santoso, Nucl.
Phys. B \textbf{606} (2001) 59.

\bibitem{stopco}C. Boehm, A. Djouadi, and M. Drees, Phys. Rev. D \textbf{62} (2000)
035012; J. R. Ellis, K. A. Olive, and
Y. Santoso, Astropart. Phys. \textbf{18} (2003) 395; J. Edsjo \etal, JCAP 0304
(2003) 001.

\bibitem{Afunnel}M. Drees and M. M. Nojiri, Phys. Rev. D \textbf{47} (1993) 376; H.
Baer and M. Brhlik, Phys. Rev. D \textbf{57}
(1998) 567; H. Baer, M. Brhlik, M. A. Diaz, J. Ferrandis, P. Mercadante, P.
Quintana, and X. Tata,
Phys. Rev. D \textbf{63} (2000) 015007; J. Ellis, T. Falk, G. Ganis, K. Olive, and
M. Srednicki, Phys.
Lett. B \textbf{510} (2001) 236; L. Roszkowski, R. Ruiz de Austri, and T. Nihei, J. High Energy Phys.
\textbf{0108} (2001) 024; A. Djouadi, M. Drees, and J. L. Kneur, J. High Energy
Phys. \textbf{0108} (2001)
055; A. Lahanas and V. Spanos, Eur. Phys. J. C \textbf{23} (2002) 185.

\bibitem{lighthiggs}P. Nath and R. Arnowitt, Phys. Rev. Lett. \textbf{70} (1993)
3696; H. Baer and M. Brhlik, Phys. Rev. D \textbf{53} (1996) 597; 
A. Djouadi, M. Drees, and J. Kneur, Phys. Lett. B \textbf{624} (2005) 60.

\bibitem{focuspoint}K. L. Chan, U. Chattopadhyay, and P. Nath, Phys. Rev. D
\textbf{58} (1998) 096004; J. L. Feng, K. T. Matchev,
and T. Moroi, Phys. Rev. Lett. 84 (2000) 2322 and Phys. Rev. D \textbf{61} (2000)
075005; see also
H. Baer, C. H. Chen, F. Paige, and X. Tata, Phys. Rev. D \textbf{52} (1995) 2746 and
Phys. Rev.
D \textbf{53} (1996) 6241; H. Baer, C. H. Chen, M. Drees, F. Paige, and X. Tata,
Phys. Rev. D \textbf{59}
(1999) 055014.

\bibitem{directDetection}CDMS Collaboration, http://cdms.berkeley.edu/; XENON100 Collaboration, http://xenon.physics.rice.edu/; Edelweiss Collaboration, http://edelweiss.in2p3.fr/; LUX Collaboration, http://lux.brown.edu/; KIMS Collaboration, http://q2c.snu.ac.kr/KIMS/KIMS\_index.htm. 


%\bibitem{EW}K.~Inoue, A.~Kakuto, H.~Komatsu, and S.~Takeshita,
%Prog.\ Theor.\ Phys.\  \textbf{68} (1982) 927;
%L.~E.~Ibanez and G.~G.~Ross,
%  Phys.\ Lett.\ B \textbf{110} (1982) 215;
%J.~R.~Ellis, D.~V.~Nanopoulos, and K.~Tamvakis,
%  %``Grand Unification In Simple Supergravity,''
%  Phys.\ Lett.\  B \textbf{121} (1983) 123.

\bibitem{isajet}
F. E. Paige \etal,
% , S.~D.~Protopopescu, H.~Baer and X.~Tata,
  %``ISAJET 7.69: A Monte Carlo event generator for p p, anti-p p, and e+ e-
  %reactions,''
  [arXiv:hep-ph/0312045].
  %%CITATION = HEP-PH/0312045;%%
 We use \isajet version 7.74.

\bibitem{higgs1}
ALEPH, DELPHI, L3, OPAL Collaborations,
G. Abbiendi \etal\ 
(The LEP Working Group for Higgs Boson Searches),
 Phys. Lett. B \textbf{565} (2003) {61}.

\bibitem{aleph}
Particle Data Group,
S. Eidelman \etal, Phys. Lett. B \textbf{592} (2004) {1}.

\bibitem{tevatron}
S. P. Das, A. Datta, and M. Maity,
Phys. Lett. B \textbf{596} (2004) 293.

\bibitem{pythia}
T. Sjostrand, S. Mrenna, and P. Skands,
% "PYTHIA 6.4 Physics and Manual."
  J. High Energy Phys. \textbf{05} (2006) 026.
  %hep-ph/0603175
  We use \pythia version 6.411 with TAUOLA.

\bibitem{pgs}
\pgs\ is a parameterized detector simulator.
We use version 4
(\url{http://www.physics.ucdavis.edu/~conway/research/software/pgs/pgs4-general.htm})
in the CMS detector configuration.
%to find gluon/light quark jets and $b$ jets.
We assume the $\tau$ identification efficiency
with $\ptvis > 20~\gev$ is 50\%, while
the probability for a jet being mis-identified
as a $\tau$ is  1\%.
%A parametrized $b$-tagging efficiency is used
%based on the CDF SECVTX $b$-tagger in
%A. Abulencia \etal, Phys. Rev. Lett. 97, 082004 (2006).
The $b$-jet tagging efficiency in PGS is
$\sim$42\% for $\et >$ 50 GeV
and $|\eta| < 1.0$, and degrading between  $1.0 < |\eta| < 1.5$.
The $b$-tagging fake rate for $c$ and light quarks/gluons is
$\sim9\%$ and $2\%$, respectively.

\bibitem{hinch1} I. Hinchliffe, F. E. Paige, M. D. Shapiro, J. Soderqvist, and W. Yao,
 %``Precision SUSY measurements at LHC,''
  Phys. Rev. D \textbf{55} (1997) 5520;
%  [arXiv:hep-ph/9610544].
  %%CITATION = PHRVA,D55,5520;%%
I. Hinchliffe and F. E. Paige,
%``Measurements in SUGRA models with large tan(beta) at LHC,''
  Phys. Rev. D \textbf{61} (2000) 095011;
 % [arXiv:hep-ph/9907519].
  %%CITATION = PHRVA,D61,095011;%%
H. Bachacou, Ian Hinchliffe, and Frank E. Paige
%``Measurements of masses in SUGRA models at LHC,''
  Phys. Rev. D \textbf{62} (2000) 015009.
 % [arXiv:hep-ph/9907518].
  %%CITATION = PHRVA,D62,015009;%%

\bibitem{massRelation}
K. Kawagoe, M. M. Nojiri, and G. Polesello
  Phys. Rev. D \textbf{71} (2005) 035008.


\bibitem{mt2}
C. G. Lester and D.J. Summers, 
Phys. Lett. B \textbf{463} (1999) {99};
W. S. Cho, K. Choi, Y. G. Kim, C. B. Park
Phys. Rev. Lett. \textbf{100} (2008) {171801}.

\bibitem{bestOriginal}
ATLAS Collaboration, N. Ozturk, [arXiv:hep-ph/0710.4546].

\bibitem{darksusy}
P. Gondolo \etal,
% J.~Edsjo, P.~Ullio, L.~Bergstrom, M.~Schelke and E.~A.~Baltz,
  %``DarkSUSY: A numerical package for supersymmetric dark matter
  %calculations,''
  [arXiv:astro-ph/0211238].
  %%CITATION = ASTRO-PH/0211238;%%

\bibitem{uncertainties}
The systematic uncertainties will be evaluated correctly once the LHC turns on.  However, for this study, we assume a $\pm3\%$ uncertainty on the energy scale for jets, taus, and missing transverse energy to estimate the systematic uncertainties independently of luminosity.  Theoretical uncertainties were not estimated in this analysis in order to see how well the dark matter relic density can be determined at the LHC.




\end{thebibliography}
\end{document}